\documentclass[aps,prl,twocolumn,showpacs,superscriptaddress]{revtex4}
\usepackage{epsfig}
\usepackage{graphicx}
\usepackage{dcolumn}
\usepackage{amsmath}

\begin{document}

\preprint{HEP/123-qed}

\title[Long Title]{Migration of nuclear shell gaps studied in
the d($^{24}$Ne,p$\gamma$)$^{25}$Ne reaction}


\author{W.N. Catford}
\affiliation{Department of Physics, University of Surrey, Guildford GU2 5XH, UK}
\author{C.N. Timis}
\affiliation{Department of Physics, University of Surrey, Guildford GU2 5XH, UK}
\author{R.C. Lemmon}
\affiliation{Nuclear Physics Group, STFC Daresbury Laboratory, Daresbury, Warrington WA4 4AD, UK}
\author{M. Labiche}
\affiliation{School of Engineering and Science, University of the West of Scotland, Paisley PA1 2BE, UK}
\affiliation{Nuclear Physics Group, STFC Daresbury Laboratory, Daresbury, Warrington WA4 4AD, UK}
\author{N.A. Orr}
\affiliation{LPC -- ENSICAEN, IN2P3/CNRS et Universit\'e de Caen, 14050 Caen, France}
\author{B. Fern\'andez-Dom\'inguez}
\affiliation{Oliver Lodge Laboratory, University of Liverpool, Liverpool L69 7ZE, UK}
\author{R. Chapman}
\affiliation{School of Engineering and Science, University of the West of Scotland, Paisley PA1 2BE, UK}
\author{M. Freer}
\affiliation{School of Physics and Astronomy, University of Birmingham, Birmingham B15 2TT, UK}
\author{M. Chartier}
\affiliation{Oliver Lodge Laboratory, University of Liverpool, Liverpool L69 7ZE, UK}
\author{H. Savajols}
\affiliation{GANIL, BP 55027, 14076 Caen Cedex 5, France}
\author{M. Rejmund}
\affiliation{GANIL, BP 55027, 14076 Caen Cedex 5, France}
\author{N.L. Achouri}
\affiliation{LPC -- ENSICAEN, IN2P3/CNRS et Universit\'e de Caen, 14050 Caen, France}
\author{N. Amzal}
\affiliation{School of Engineering and Science, University of the West of Scotland, Paisley PA1 2BE, UK}
\author{N.I. Ashwood}
\affiliation{School of Physics and Astronomy, University of Birmingham, Birmingham B15 2TT, UK}
\author{T.D. Baldwin}
\affiliation{Department of Physics, University of Surrey, Guildford GU2 5XH, UK}
\author{M. Burns}
\affiliation{School of Engineering and Science, University of the West of Scotland, Paisley PA1 2BE, UK}
\author{L. Caballero}
\affiliation{Instituto de Fisica Corpuscular, CSIC-Universidad de Valencia, 46071 Valencia, Spain }
\author{J.M. Casadjian}
\affiliation{GANIL, BP 55027, 14076 Caen Cedex 5, France}
\affiliation{IRFU, CEA-Saclay, 91191 Gif-sur-Yvette, France }
\author{N. Curtis}
\affiliation{School of Physics and Astronomy, University of Birmingham, Birmingham B15 2TT, UK}
\author{G. de France}
\affiliation{GANIL, BP 55027, 14076 Caen Cedex 5, France}
\author{W. Gelletly}
\affiliation{Department of Physics, University of Surrey, Guildford GU2 5XH, UK}
\author{X. Liang}
\affiliation{School of Engineering and Science, University of the West of Scotland, Paisley PA1 2BE, UK}
\author{S.D. Pain}
\affiliation{Department of Physics, University of Surrey, Guildford GU2 5XH, UK}
\author{V.P.E. Pucknell}
\affiliation{Nuclear Physics Group, STFC Daresbury Laboratory, Daresbury, Warrington WA4 4AD, UK}
\author{B. Rubio}
\affiliation{Instituto de Fisica Corpuscular, CSIC-Universidad de Valencia, 46071 Valencia, Spain }
\author{O. Sorlin}
\affiliation{GANIL, BP 55027, 14076 Caen Cedex 5, France}
\author{K. Spohr}
\affiliation{School of Engineering and Science, University of the West of Scotland, Paisley PA1 2BE, UK}
\author{Ch. Theisen}
\affiliation{IRFU, CEA-Saclay, 91191 Gif-sur-Yvette, France }
\author{D.D. Warner}
\affiliation{Nuclear Physics Group, STFC Daresbury Laboratory, Daresbury, Warrington WA4 4AD, UK}

\date{\today}

\begin{abstract}
The transfer of neutrons onto $^{24}$Ne has been measured using a reaccelerated radioactive beam of $^{24}$Ne to study the (d,p) reaction in inverse kinematics. The unusual raising of the first 3/2$^+$ level in $^{25}$Ne and its significance in terms of the migration of the neutron magic number from N=20 to N=16 is put on a firm footing by confirmation of this state's identity. The raised 3/2$^+$ level is observed simultaneously with the intruder negative parity 7/2$^-$ and 3/2$^-$ levels, providing evidence for the reduction in the N=20 gap. The coincident gamma-ray decays allowed the assignment of spins as well as the transferred orbital angular momentum. The excitation energy of the 3/2$^+$ state shows that the established USD shell model breaks down  well within the {\it sd} model space and requires a revised treatment of the proton-neutron monopole interaction.
\end{abstract}


\maketitle

The monopole part of the nucleon-nucleon interaction is now recognised as having a major effect on nuclear shell structure far from stability \cite{otsuka01,sorlin-review}. The interaction between valence protons and neutrons is sufficient to alter the energies of single-particle levels so that different magic numbers (or shell gaps) appear and this can substantially affect the collective \cite{sorlin-cr64} and magnetic \cite{neyens-mg33} properties and basic quantities such as the lifetime \cite{s44}. Nucleon transfer reactions induced by light ions are an established experimental tool for studying single-particle structure \cite{lee}.  Here we employ the (d,p) reaction in inverse kinematics to explore the disappearance of the N=20 magic number (and its replacement by N=16) in the neutron-rich neon isotones. As will be shown, the measurement of the differential cross sections of the light ejectiles plus the coincident gamma decays of the residual nucleus brings a new power to this type of study.

Recent work using other techniques has provided evidence for the emergence of N=16 as a magic number in this region, but has not identified the single-particle structure in an unambiguous manner through measurements of the spectroscopic factors and spins. In a study of the $\beta$-decay of $^{25}$F \cite{padgett05} the increased energy of the $0d_{3/2}$ neutron orbital was inferred. This made use of a preliminary analysis of the present work \cite{catford05a}, and concluded that the energy shift was consistent with the monopole effect \cite{padgett05}. In a study of $^{27}$Ne using the (d,p$\gamma$) reaction but without detecting the protons \cite{obertelli06}, a reduced gap between the $0d_{3/2}$ and higher negative parity orbitals was deduced. This agreed with nucleon removal studies \cite{terry04}. Finally, in recent studies of $^{23}$O by transfer \cite{elekes07} and $^{25}$O by proton removal \cite{hoffman08} the $0d_{3/2}$ state was found to have an increased excitation energy but the required modifications to the shell model interaction were not mutually consistent \cite{elekes07,hoffman08}. While an extensive review including the emergence of the N=16 magic number has recently been published \cite{sorlin-review}, further quantitative data are needed in order to understand this monopole effect properly.

In this spirit, we have investigated the levels of $^{25}$Ne, in which the valence neutron can occupy the $1s_{1/2}$ or $0d_{3/2}$ orbitals, or the higher lying negative parity orbitals ($0f_{7/2}$, $1p_{3/2}$, $\ldots$). This explores the N=16, N=20 and N=28 gaps. Existing information about levels in $^{25}$Ne is included in Table~I. The $\beta$-decay work showed \cite{padgett05}, in addition to the two excited states listed, three other positive parity states above 3.3 MeV including a state seen also in nucleon removal from $^{26}$Ne and proposed to have $J^\pi = 5/2^+$ \cite{terry04}. The present 3.33 MeV state was proposed \cite{woods85} to have negative parity (along with the 4.07 MeV state) based on the ($^{13}$C,$^{14}$O) selectivity.

A pure beam of $^{24}$Ne ions, at 10.6\,$\cdot$A MeV and $2 \times 10^5$ particles/sec, was provided by the SPIRAL facility at GANIL. The beam impinged on a self-supporting target of deuterated polythene (CD)$_2$ with  thickness 1 mg/cm$^2$. The target was surrounded by TIARA \cite{labiche05, catford05b}, an array of silicon detectors spanning 85\% of 4$\pi$. The TIARA coverage from 36$^\circ$ to 144$^\circ$ comprised an octagonal ``barrel'' of position sensitive (resistive strip) detectors, 400$\mu$m thick. The most backward angles (144$^\circ$ to 169$^\circ$) were covered by an array of 400$\mu$m thick double-sided Si strip detectors (DSSSDs). In very close proximity to the target (50mm) the front faces of four EXOGAM segmented Ge detectors \cite{simpson00} formed four sides of a cube spanning 66\% of 4$\pi$. The full-energy peak efficiency for $\gamma$-ray detection was estimated to be 15\% at 1.0 MeV. During the experiment, however, the Ge detectors were in operation for only a fraction of the beam exposure. Beam particles and beam-like reaction products emerging near zero degrees were analysed by the VAMOS magnetic spectrometer \cite{savajols99}. The focal plane detectors gave Z identification by $\Delta$E-E and mass by time-of-flight \cite{catford05b}. The start signal for the timing was provided by a thin (10$\mu$m) plastic scintillator mounted 0.5m upstream of the reaction target.  The data acquisition was triggered if an event was recorded in any of the TIARA silicon detectors. Any coincident event in VAMOS was recorded and, in that case, also any further coincident event in EXOGAM. Only the core signals \cite{simpson00} from EXOGAM were available, limiting the accuracy of the Doppler energy corrections for $\gamma$-rays emitted by the beam-like particles ($\beta \approx 0.1$) to a FWHM of 65 keV at 1 MeV.

Products of $^{24}$Ne reacting with $^{12}$C in the target were evident in the TIARA singles spectra of energy against angle, comprising knock-on $^{12}$C plus protons and $\alpha$-particles from fusion-evaporation, but were eliminated by requiring a beam or beam-like particle in VAMOS. The elastically scattered deuterons were detected just forward of 90$^\circ$ \cite{catford02} in coincidence with scattered beam recorded in VAMOS. These measurements provided an absolute normalization of all the differential cross sections. Protons from the (d,p) reaction were recorded from the most backward angles until the increasing energies resulted in their penetrating the barrel near 95$^\circ$. The protons corresponding to the most forward c.m.\ angles were detected in the DSSSDs between 135$^\circ$ and 170$^\circ$, with improved energy resolution and lower energy thresholds compared to the resistive strip detectors in the barrel.

\begin{figure}[ht]
\begin{center}
\psfig{figure=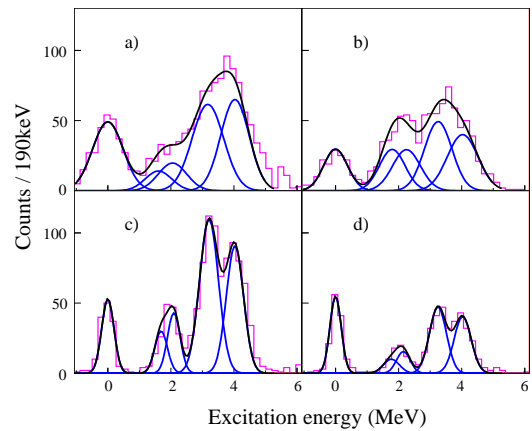,width=0.8\columnwidth,angle=0}
\end{center}
\caption{Excitation energy spectra deduced from protons detected at laboratory angles of: (a) 109$^\circ$, (b) 125$^\circ$, (c) 149$^\circ$, (d) 164$^\circ$ (color online).}
\label{ex-spec}
\end{figure}

\begin{figure}[ht]
\begin{center}
\psfig{figure=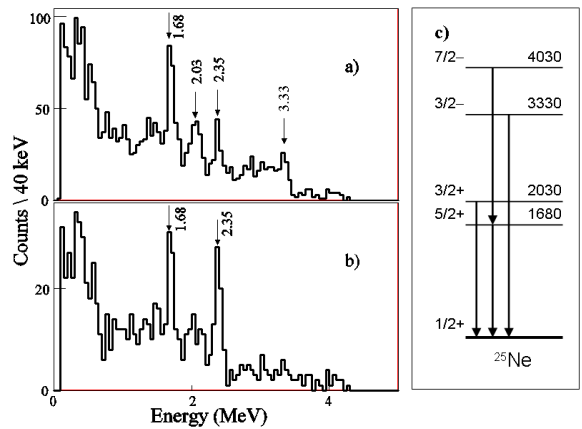,width=0.9\columnwidth,angle=0}
\end{center}
\caption{Doppler-corrected gamma-ray energy spectra recorded in coincidence with (a) all protons and (b) protons populating the state in $^{25}$Ne at 4.03 MeV; (c) levels in $^{25}$Ne.}
\label{gamma}
\end{figure}

The particles observed in TIARA in coincidence with $^{25}$Ne were used to calculate the $^{25}$Ne excitation energy, according to the measured energies and angles and the kinematics for (d,p). The resolution in excitation energy was strongly angle dependent, owing to the target effects and the detector resolution \cite{winfield97}. In total, 8 angular bins were used, including 5 in the barrel. Fig. \ref{ex-spec} shows example excitation energy spectra. These were fitted with a series of Gaussian peaks, using the $\gamma$-ray data to identify the peaks to be included and their energies. The $\gamma$-ray energy spectrum from all triple coincidence $^{25}$Ne-p-$\gamma$ events, after Doppler correction, is displayed in Fig. \ref{gamma}(a). Gamma-ray peaks were observed at energies of 1.68, 2.03, 2.35 and 3.33 MeV ($\pm$ 0.04 MeV). As discussed below, the $\gamma$-ray at 2.35 MeV is in cascade with that at 1.68 MeV, suggesting a state at 4.03 MeV. With these 4 excitation energies and the ground state fixed, the spectra for the various angle bins were each fitted with 5 peaks of equal (but angle-dependent) width.

\begin{figure}[hb]
\begin{center}
\psfig{figure=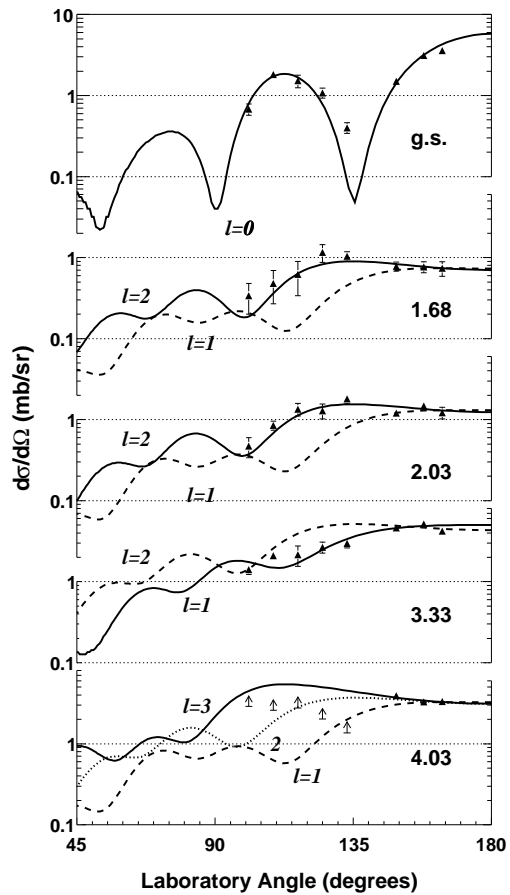,width=0.85\columnwidth,angle=0}
\end{center}
\caption{Differential cross sections for final states in $^{25}$Ne, labeled by excitation energy and compared with normalized ADWA calculations. The solid line for each state corresponds to the adopted $\ell$-transfer and spin as included in Table~I. For the points indicating lower limits (arrows), the detector energy thresholds reduced the efficiency.}
\label{angdis}
\end{figure}

The angular distributions from the (d,p) reaction populating states in $^{25}$Ne are shown in Fig. \ref{angdis}, together with calculations using the Adiabatic Distorted Wave Approximation (ADWA) \cite{johnson70} with standard parameters that have been shown to work well for comparisons with large-basis shell-model calculations \cite{lee}. The nucleon-nucleus potential of Bechetti-Greenlees was employed \cite{bechetti}.  The elastic (d,d) cross section was measured \cite{catford05b} between 25$^\circ$ and 45$^\circ$ (c.m.) simultaneously with the (d,p) and fitted using optical potential parameters obtained from d+$^{26}$Mg at 6\,$\cdot$A MeV \cite{meurders74}. This calibrated the product of beam exposure and target thickness.

The angular distribution for transfer to the $^{25}$Ne ground state exhibits a characteristic dip near 135$^\circ$ (17$^\circ$ in the c.m.\ frame) indicating an $\ell = 0$ transfer. The angular distributions of the 1.68 MeV and 2.03 MeV states are consistent only with $\ell = 2$ transfer and are clearly different from that of the ground state. The other strong peaks at 3.33 and 4.03 MeV are in a region where USD shell-model calculations
(such as those included in Table~I) predict little transfer strength to positive parity states. For the 3.33 MeV state, the fall-off in yield with increasing c.m.\ angle (decreasing laboratory angle) is too rapid for 
high $\ell$-transfers and implies $\ell = 1$. For the 4.03 MeV state, the experiment provided only a lower limit on the cross section at some angles. This resulted from the loss of some counts below the energy threshold of the barrel detectors, which is increasingly important near the end of the barrel (135$^\circ$) owing to the kinematics. The observed yield near 90$^\circ$ is sufficient, however, to rule out any rapid fall-off with increasing c.m.\ angle, and in fact the data are consistent only with an $\ell$-transfer of at least $\ell = 3$, with $\ell = 2$ giving too low a yield near 90$^\circ$ compared to that near 170$^\circ$. Since there are no $\ell = 4$ orbitals nearby in energy in the shell model, an $\ell = 3$ transfer is deduced for the 4.03 MeV state.

\begin{table*}[th]
\caption{Results for states in $^{25}$Ne. The measured excitation energies are compared with previous work, where the experimental errors are 40~keV (present), of order 1~keV \cite{reed99,padgett05}, 30~keV \cite{woods85} and 50-80~keV \cite{wilcox73}. States observed in the beta-decay work but not expected in transfer have been omitted, as have states at 4.7$\pm$0.1 MeV \cite{wilcox73} and 6.28$\pm$0.05 MeV \cite{woods85}. }
\begin{tabular}{ccccccccccccccc}
\hline
{} &{} &{} &{} &{}\\[-1.5ex]
E$_{\rm x}$(keV) & E$_{\rm x}$(keV)& E$_{\rm x}$(keV)& E$_{\rm x}$(keV)& E$_{\rm x}$(keV) & $\ell$ & $J^\pi$ & $S$ & E$_{\rm x}$ (keV) &  $S$  & E$_{\rm x}$(keV) &  $S$ & E$_{\rm x}$(keV) &  $S$\\
(d,p$\gamma$) & $\beta $-$\gamma$  & $\beta $-$\gamma$  &  ($^{13}$C,$^{14}$O) & ($^7$Li,$^8$B) & (~$\hbar$~) & & ~ & USD & USD & USD-A & USD-A & USD-B & USD-B\\
present &  \cite{reed99} &   \cite{padgett05} &  \cite{woods85} &  \cite{wilcox73} & present & & present &  \cite{brownsde} & & \cite{brown-new} & ~ & \cite{brown-new} & ~ \\[1ex]
\hline
{} &{} &{} &{} &{} &{}\\[-1.5ex]
0 & 0  & 0 & 0 & 0  & 0 &  1/2$^+$ & 0.80 & 0  & 0.63  & 0 & 0.64 & 0 & 0.64\\[1ex]
1680 & 1703 & 1702 & 1740 & 1650  & 2 & 5/2$^+$ & 0.15 & 1779  & 0.10 & 1850 & 0.10 & 1756 & 0.10 \\[1ex]
2030 & - & 2090 & - & 2030 & 2 & 3/2$^+$ & 0.44 & 1687  & 0.49 & 2042 & 0.42 & 2043 & 0.39 \\[1ex]
3330 & - & - & 3330 & 3250 & 1 & 3/2$^-$ & 0.75 &  &  &  \\[1ex]
4030 & - & - & 4070 & 4050 & 3 & 7/2$^-$ & 0.73 &  &  &  \\[1ex]
\hline
\end{tabular}\label{specfacs} \\
\vspace*{-13pt}
\end{table*}

The deduced $\ell$-transfers can be extended to spin assignments using the observed $\gamma$-ray decay scheme. The $\gamma$-ray spectrum obtained using a restricted gate on excitation energy near 4 MeV is shown in Fig. \ref{gamma}(b). When the quadruple-coincidence $^{25}$Ne-p-$\gamma$-$\gamma$ data are analyzed, the limited statistics are sufficient to demonstrate that a gate on the 1.68 MeV peak highlights the peak at 2.35 MeV and, similarly, a 2.35 MeV gate highlights the 1.68 MeV transition \cite{catford05b}. Such a cascade is of course expected for a (5/2$^-$, 7/2$^-$) $\ell=3$ state, decaying via an $\ell=2$ state, and the cascade from a 7/2$^-$ state would select just the 5/2$^+$ state. From simple shell model considerations, the lowest $\ell=3$ state is expected to be 7/2$^-$. Indeed, the isotonic reaction $^{26}$Mg(d,p)$^{27}$Mg \cite{meurders74} strongly populates an $\ell=3$ state at 3.76 MeV which is assigned 7/2$^-$ and decays via the 5/2$^+$ state \cite{endt78}. The combined evidence implies that the 4.03 MeV state in $^{25}$Ne should be assigned 7/2$^-$ and also that the intermediate 1.68 MeV state in the decay is 5/2$^+$. Furthermore, the reaction yield to the 1.68 MeV state is lower than that for the 2.03 MeV state in $^{25}$Ne. This reinforces the 5/2$^+$ assignment, since the 5/2$^+$ hole state (naively $\nu (1s_{1/2}^2 \otimes 0d_{5/2}^{-1})$) should be populated more weakly than the 3/2$^+$ particle state ($\nu (0d_{3/2})$), as also observed in $^{27}$Mg \cite{meurders74}. By elimination, the $\ell=2$ state at 2.03 MeV can be inferred to be the 3/2$^+$ state expected from systematics and shell model considerations. These arguments for the $\ell=2$ states are all supported by high-energy nucleon removal \cite{terry04} and $\beta$-decay \cite{padgett05} studies. The $^9$Be($^{26}$Ne,$^{25}$Ne) reaction was seen to populate the 5/2$^+$ hole state but not the 3/2$^+$ state \cite{terry04}. The $^{25}$F $\beta$-decay was found to favour the 2090 keV state relative to that at 1702 keV and comparison with USD calculations suggested that the lower state must be 5/2$^+$, leaving the higher energy state as 3/2$^+$ \cite{padgett05}. Finally, the lowest energy $\ell=1$ state is expected to be 3/2$^-$ in the shell model, and hence the 3.33 MeV state in $^{25}$Ne will correspond to the strongly populated 3.56 MeV 3/2$^-$ state in $^{27}$Mg. As expected, this state in $^{25}$Ne $\gamma$-decays to the 1/2$^+$ ground state.

The excitation energies and deduced spin-parity assignments are listed in Table~I. The adopted energies for the first two excited states are defined to 1 keV precision by the beta-decay results \cite{padgett05} whilst the weighted average energies for the negative parity states are $3325 \pm 25$ keV and $4055 \pm 25$ keV. The precision in the present work is 40 keV owing to the limited statistics and the large widths of the gamma-ray peaks. The relative ordering of the lowest 7/2$^-$ and 3/2$^-$ levels is the same as in the isotone $^{27}$Mg, but they are further apart in $^{25}$Ne. This contradicts Monte-Carlo SDPF-M shell model predictions for $^{25}$Ne \cite{terry04}. Much more dramatically, whilst the 5/2$^+$ state is at a similar energy in the two nuclei (reflecting in part the separation of the $0d_{5/2}$ and $1s_{1/2}$ orbitals), the lowest 3/2$^+$ state moves from about 700 keV below the 5/2$^+$ in $^{27}$Mg to 250 keV above in $^{25}$Ne. This is unexpected according to standard USD shell-model calculations \cite{brownsde} (Table~I) and reflects a raising in energy of the $0d_{3/2}$ orbital in neutron-rich nuclei. This can be reproduced with a revised treatment of the monopole interaction \cite{padgett05,brown-new}. The present data show that the raising of the $0d_{3/2}$ orbital simultaneously opens the N=16 gap and closes the N=20 gap as the negative parity states are at similar energies in $^{27}$Mg and $^{25}$Ne.

Having adopted the $J^\pi$ assignments listed in Table~I, the ADWA calculations can be scaled to the experimental data and spectroscopic factors ($S$) deduced. These are also listed in Table~I, where a value of unity would indicate transfer to an unblocked pure single-particle orbital. The uncertainty arising from the theoretical model is estimated as 20\% \cite{lee}. The 1/2$^+$, 3/2$^-$ and 7/2$^-$ states all carry a substantial part of the single-particle strength. The 5/2$^+$ spectroscopic factor is smaller because it is predominantly a hole state and as such effectively measures the amount of $\nu (0d_{5/2})^2$ to $\nu (1s_{1/2})^2$ pair excitation in $^{24}$Ne. The 3/2$^+$ state is evidently of complex structure; besides the $0d_{3/2}$ single-particle structure it includes, for example, a component of $^{24}$Ne($2^+$) $\otimes$ $1s_{1/2}$. According to USD calculations, the remaining $0d_{3/2}$ single-particle strength is distributed widely, with much in the third and fifth 3/2$^+$ states between 4.5 and 6 MeV. We note that in the N=15 isotone $^{23}$O, the 3/2$^+$ state was observed \cite{elekes07} (with $S$~=~0.5$\pm$0.1) at an excitation energy well above that predicted by the USD interaction \cite{brownsde}.  Interestingly, while this could be reproduced using the revised USD interactions \cite{brown-new}, they fail to explain the excitation energy of the corresponding state in $^{25}$O \cite{hoffman08}.

In conclusion, the spins of the low-lying levels in $^{25}$Ne have been firmly assigned. The inversion of the 3/2$^+$ and 5/2$^+$ levels represents a significant breakdown of the USD model, which can be explained in terms of a monopole shift in the effective $0d_{3/2}$ single-particle energy \cite{padgett05,brown-new} so as to open a shell gap at N=16. The measured spectroscopic factors for $^{25}$Ne provide the means to test in detail the shell model modifications invoked to explain this shift. The present data also indicate that the opening of the N=16 gap is accompanied by a closing of the N=20 gap. This confirms the inference from $^{27}$Ne studies \cite{obertelli06,terry04}. Finally, the experimental methods of (d,p$\gamma$) and (d,p$\gamma \gamma$) employed here are likely to find wide applicability in the exploration of nuclear structure with radioactive beams.

The authors acknowledge the excellent support provided by the technical staff of LPC and GANIL.

\end{document}